# Social Technologies for Developing Collective Intelligence in Networked Society

PROF.DR.AELITA SKARŽAUSKIEN , Mykolas Romeris University, Lithuania
PROF.DR.BIRUT PITR NAIT -ŽIL NIEN , Mykolas Romeris University, Lithuania
DR.EDGARAS LEICHTERIS, Knowledge Economy Forum, Lithuania
DR.ŽANETA PAUKSNIEN , Mykolas Romeris University, Lithuania

1. INTRODUCTION

Europe 2020 strategy and Digital Agenda for EU sets political framework to achieve smart and inclusive European development based on ICT technologies. Surveys conducted by analysts such as Forrester Research [2009] and McKinsey Global institute [2012] demonstrate that social technologies (ST) continue to grow in popularity inside the society and these developments will have an influence on policies and drive economic and societal changes. Social media has exploded, transforming the way that people share and consume information. "Social networks are becoming the preferred method of communication for new generations and communication styles are evolving into a more collaborative approach" [Alberghini, et al. 2010]. Following the Internet design societies, organizations and movements have evolved from bureaucratic/centralized to both decentralize and distributed networks [Barahona et al 2012]. A network is a formal structure, where intelligent activities emerge [Monge and Contractor 2003]. The recent successes of systems like Google, Wikipedia or InnoCentive suggest that individuals and groups can more effectively create valuable intellectual products by acting on the basis of a collective intelligence. The goal of the research is to understand how to take advantage of these possibilities to tackle societal challenges. Since the future is basically unpredictable and uncertain, society must rely on creative initiatives from the citizens to be able to create the desired future [Johannessen et al 2001]. The scientific problem in our project is defined as a question: how social technologies could contribute to the development of smart and inclusive society? The subject of our research are networked projects (virtual CI systems) which include collective decision making tools and innovation mechanisms allowing and encouraging individual and team creativity, entrepreneurship, online collaboration, new forms of self-regulation and self-governance, self-configuration of communities by considering these projects as being catalyst for emergence of CI. The answers to these theoretical questions could have huge practical implications by influencing more reasonable and sophisticated application of social technologies in practise [Skarzauskiene, Pitrenaite 2013].

1.1 The Context for Development of Networked Community Projects in Lithuania

One of the flagship initiatives of Digital agenda for Europe aims to create single digital market based on fast/ultrafast internet and interoperable applications. As the response to that challenge Lithuania started RAIN I and RAIN II projects carried out by absorbing EU structural support. Owing to Rural Internet Access Points (RIAPs) the fast and high-quality internet became accessible not only in cities but also to rural areas public sector, business organizations and residents. It is planned that by the end of 2014 broadband internet will reach 98.7 percent of rural areas in Lithuania. There is no doubt that the widespread and availability of the internet is one of the prerequisites for new form of interconnection, different forms of social cohesion and conditions to collectively build community interactions. "High-speed broadband has the potential to fundamentally alter communication practices within the community … influence transformation of culture and society" [Institute for a Broadband- Enabled Society 2013].
The Web's growth in reach and capability set the stage for the explosive growth of networked projects in Lithuania, funded by public organizations or private entities. Among them are such projects as

*manobalsas.lt* (My Voice Lt, www.manobalsas.lt), *manoseimas.lt* (My Parliament, www.manoseimas.lt), eVoting testing system *ivote.lt* (www.ivote.lt), Lithuanian civic initiative think tank *Aš Lietuvai.lt* (*I for Lithuania*, www.aslietuvai.org), the platform for e-democracy *Lietuva2.0.lt* (*Lithuania2.0*, www.lietuva2.lt) etc. High quality IT infrastructure and boost of virtual social projects could become a possibility to effect positive changes in communities and government. However, Lithuanian society encounters a social challenge – Lithuania remains a state where civic engagement for many socio-cultural reasons and post-soviet mentality is poor. The researches that are being conducted by Civil Society Institute since 2007 exhibit low level of the society's political self-awareness – in 2012 the civic engagement was rated in average 38,4 of possible 100 points [Civil Society Institute, 2013]. Since 2007 this rate is increasing very slightly. It is worth mentioning that civic engagement of young people (from 15 to 29 years old) is distinguished as being significantly low. The social environment for civic engagement in Lithuania is evaluated only 22,2 of 100 points in 2012 and is revealed to be adverse. 6–7 out of 10 individuals have negative opinion on participation environment. That could be one of the reasons why society's general interest in public issues remains only average (evaluated about 40 points of 100 during the last 3 years).

Striving to explore the extent and major trends of the engagement and participation of Lithuanian society in virtual community projects as platforms for emergence of CI, we conducted a quantitative survey. A public opinion and market research company executed the survey in November – December 2013. Statistically reliable sample (with the confidence level of 95 percent) of Lithuanian population was composed of 1022 respondents applying random stratified selection. The sample included men and women with age ranging from 15 to 74 from cities and rural areas of all 10 Lithuanian counties. An original research instrument was designed and included these main sections: I – the level of interest in social technologies; II – the level of knowledge about the virtual community projects tackling societal problems; III – the content of the process of participation in virtual community projects; IV – the level of satisfaction in virtual communication. Results of the survey demonstrate active exploitation of internet resources by Lithuanian population as 44 percent of respondents use internet every day, other 20 percent – at least once a week. The majority of those who use internet (from 59 to 67 percent) exploit it for business and private communication, information search, entertainment, financial operations and only 21 percent put some contents for the public (for instance write comments, participate in discussions, create articles and blogs). 61 percent of frequent internet users surf different virtual communication networks. However, only 2 to 6 percent of them participate in public issues oriented virtual activities such as discussing societal problems, creating and proposing ideas for social projects, joining social projects implementation, voting for public problem solving ideas and alternative decisions. Moreover, only 7 percent of frequent internet users join virtual community projects focused on tackling societal problems, and only 1 to 5 percent of internet surfers are familiar with previously mentioned networked projects *My Parliament, I for Lithuania, Lithuania 2.0* etc. Those who do not participate in social issues oriented virtual networks, justify their passiveness by lack of interest (54 percent), lack of time (44 percent), disbelief that their opinion is valued (11 percent) and that initiatives could change the reality (16 percent). However, the majority of those who take part in these projects (66 percent) are satisfied with their activities.

The research results prove the necessity to search for tools fostering engagement of society to propose managerial, social and legal measures for the stimulation of the process. Our next step is the intent to propose a set of criteria for measuring Collective intelligence in networked platforms (virtual CI systems).

### 1.2   CI Index: Findings and Directions for Further Research

The methodology for CI index calculation allows the analysis, evaluation and assessment of significant changes in CI systems and will be based on predefined questionnaire, automatic data collection and their algorithmic analysis. The CI index will show the state and dynamics of the CI according to changes of various internal and external parameters [Spila 2012]. Virtual research environment with required software for scientific research activities will be created to be able to develop the proposed methodology and to apply it for the research activities. The data necessary for the identification of the CI Index parameters were collected during the quantitative and qualitative research and will be revised during the scientific experiment. A longitudinal observation of number of networked platforms will be undertaken to measure agreed representative parameters. Below we introduce the theoretical framework for CI Index, dimensions and proposed indicators for a scientific discussion (see Figure1).

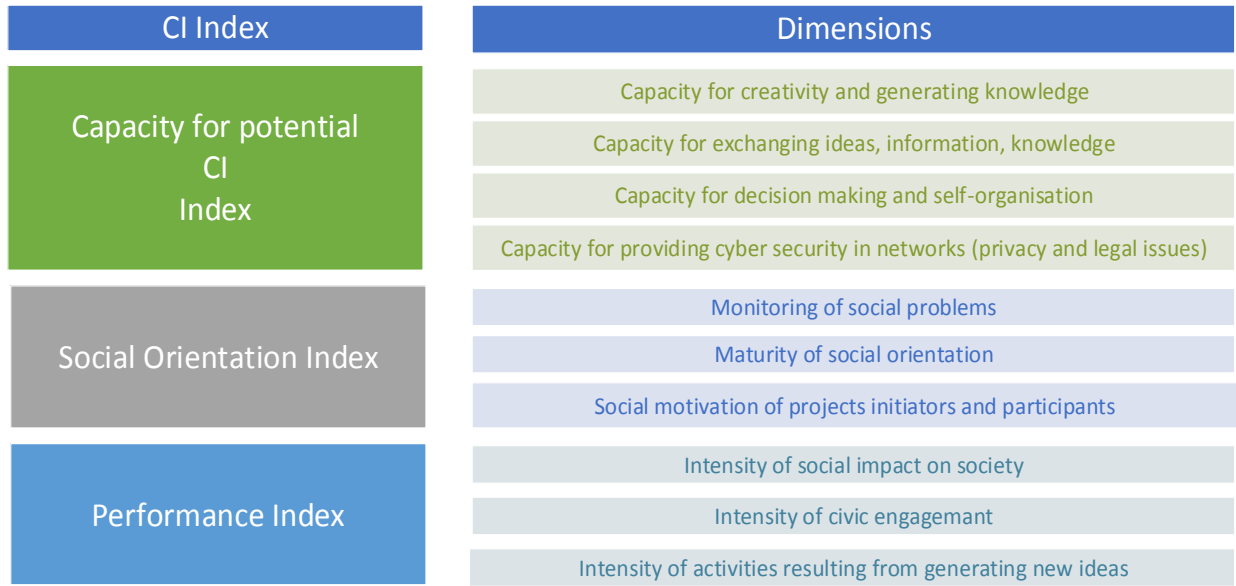

Fig.1 CI Index Model

We have used the proposed methodological framework for Social Innovation Index calculation by creating CI model [Spila 2012]. For the operational purposes the CI Index has been designed around three indices, which are defined by different dimensions. We elaborated on various dimensions which cover different aspects of each of the index and created a different group of indicators to measure each dimension, for example, the *Capacity for Creativity* includes two groups of indicators: a) platform indicators ((*Degree of participants diversity; Size of the group; Degree of motivation, involvement of group members; Degree of freedom and security to offer idea etc.*) and b) *Social Technologies indicators* (*Level of development possibilities; Degree of user friendliness; Speed and convenience; Existence of mechanism for collective brainstorming; Existence of mechanism for anonymous offering of ideas etc.*). The first group of indicators reflects from grouping of different questions in the questionnaire about platform itself, the second – questions about technological parameters.

The methodology will allow to identify and analyze conditions that lead communities to become more collective intelligent, but first following research questions needs to be answered: how to increase engagement of passive society into collective decision making process, how technologies could help to structure the information, purify the positions, reconcile different opinions and formulate the real society voice etc. To solve social challenges further research is planned in the field with the aim to show how existing social technology parameters might help platform developers to create new IT based applications fostering self-organization, collective decision making and learning etc.

ACKNOWLEDGMENTS
The research is funded by European Social Fund under the measure „Support to Research Activities of Scientists and Other Researcher"(Global Grant) administrated by Lithuanian Research Council (Grant No. VP1-3.1-ŠMM-07-K-03-030, Social technologies for Developing Collective Intelligence in Networked Society).